\documentclass[10pt, conference, letterpaper]{IEEEtran}
\usepackage{algorithmic, algorithm}
\usepackage[cmex10]{amsmath}
\usepackage{amssymb, latexsym}
\usepackage[caption=false,font=footnotesize]{subfig}
\usepackage{cite}
\usepackage{pgfplots}
\usepackage{enumitem}
\usepackage{microtype}
\usepackage{tikz}

\usetikzlibrary{patterns,decorations.pathreplacing,shapes,arrows,chains}

\pgfplotsset{compat=newest}
\pgfplotsset{plot coordinates/math parser=false}
\newlength\figureheight
\newlength\figurewidth

\definecolor{ref}{rgb}{0.65,0.65,0.65} 
\definecolor{lhmm}{rgb}{0.9,0.6,0.5}
\definecolor{ghmm}{rgb}{0.7,0.9,0.35}
\definecolor{lsvm}{rgb}{0.9,0.8,0.25}
\definecolor{gsvm}{rgb}{0.4,0.8,0.9}

\definecolor{bblue}{HTML}{4F81BD}
\definecolor{rred}{HTML}{C0504D}
\definecolor{ggreen}{HTML}{9BBB59}
\definecolor{ppurple}{HTML}{9F4C7C}
\definecolor{ggray15}{HTML}{D9D9D9}

\setlist[itemize]{wide}

\begin{document}

\title{PopNetCod: A Popularity-based Caching Policy for Network Coding enabled Named Data Networking}

\author{
\IEEEauthorblockN{Jonnahtan~Saltarin\IEEEauthorrefmark{1}, Torsten~Braun\IEEEauthorrefmark{1}, Eirina~Bourtsoulatze\IEEEauthorrefmark{2} and Nikolaos~Thomos\IEEEauthorrefmark{3}}
\IEEEauthorblockA{\IEEEauthorrefmark{1}University of Bern, Bern, Switzerland}
\IEEEauthorblockA{\IEEEauthorrefmark{2}Imperial College London, London, United Kingdom}
\IEEEauthorblockA{\IEEEauthorrefmark{3}University of Essex, Colchester, United Kingdom}
\IEEEauthorblockA{saltarinj@gmail.com, braun@inf.unibe.ch, e.bourtsoulatze@imperial.ac.uk, nthomos@essex.ac.uk}
}

\IEEEoverridecommandlockouts\IEEEpubid{\makebox[\columnwidth]{ISBN 978-3-903176-08-9~\copyright~2018 IFIP \hfill} \hspace{\columnsep}\makebox[\columnwidth]{ }}

\maketitle

\begin{abstract}
In this paper, we propose PopNetCod, a popularity-based caching policy for network coding enabled Named Data Networking. PopNetCod is a distributed caching policy, in which each router measures the local popularity of the content objects by analyzing the requests that it receives. It then uses this information to decide which Data packets to cache or evict from its content store. Since network coding is used, partial caching of content objects is supported, which facilitates the management of the content store. The routers decide the Data packets that they cache or evict in an online manner when they receive requests for Data packets. This allows the most popular Data packets to be cached closer to the network edges. The evaluation of PopNetCod shows an improved cache-hit rate compared to the widely used Leave Copy Everywhere placement policy and the Least Recently Used eviction policy. The improved cache-hit rate helps the clients to achieve higher goodput, while it also reduces the load on the source servers.
\end{abstract}

\section{Introduction}
\label{sec:introduction}

Data intensive applications, \textit{e.g.}, video streaming, software updates, \textit{etc.}, are the major sources of data traffic in the Internet, and their predominance is expected to further increase in the near future~\cite{CiscoVNI}. Moreover, nowadays Internet users are more concerned about \textit{what} data they request, rather than \textit{where} that data is located. To address the increased data traffic and the shift in interest from location to data, technologies like Content Delivery Networks (CDN) have been proposed. However, these solutions cannot fully exploit the network resources and deal effectively with the increasing amount of data traffic, since they work on top of the current Internet architecture, which is based on host-to-host communication. To address this issue, the Named Data Networking (NDN) architecture~\cite{Zhang2014,Jacobson2009} has been proposed, which replaces the addresses of the communicating hosts (\textit{i.e.}, IP addresses) with the name of the data being communicated. In the NDN architecture, clients request data by sending an \textit{Interest} that contains the name of the requested data. Any network node that receives the Interest and holds a copy of the requested data can satisfy it by sending a \textit{Data packet} back to the client.

Two of the main advantages that the NDN architecture has over the traditional host-to-host architectures are: \textit{(i)} the inherent use of in-network caching, and \textit{(ii)} the built-in support for multipath communications. The pervasive in-network caching concept proposed by NDN reduces the number of hops that Interests and Data packets need to travel in the network. This reduces the delay perceived by the application retrieving the requested data. However, having caches in all the routers is not always necessary to yield the full benefits that caching brings to the data delivery process. Previous works~\cite{Dabirmoghaddam2014, Fayazbakhsh2013, Sun2014} have shown that enabling caches only at the edge of the network may achieve performance improvements similar to those obtained when every router is equipped with a cache. Furthermore, NDN provides natural multipath support by allowing clients to distribute the Interests that they need to send to retrieve content objects over all their network interfaces (\textit{e.g.}, LTE, Wi-Fi), which enables the applications to better use the clients' network resources. However, in the presence of multiple clients and/or multiple data sources, the optimal use of multiple paths requires the nodes to coordinate where they forward each Interest in order to reduce the number of Data packet transmissions and the network load.

To optimally exploit the benefits brought by in-network caching and multipath communication, previous works~\cite{Montpetit2012,Saltarin2016} had proposed the use of network coding~\cite{Ahlswede2000}. In a network coding enabled NDN architecture, the network routers code Data packets by combining the Data packets available at their caches prior to forwarding them. The use of network coding \textit{(i)} increases Data packet diversity in the network, hence, the use of in-network caches is optimized, and \textit{(ii)} in multi-client and multi-source scenarios it removes the need for coordinating the faces where the nodes forward each Interest, which enables efficient multipath communication. Although there are works that consider the use of network coding in NDN, they do not consider that caching capacity is limited~\cite{Montpetit2012,Saltarin2016,Saltarin2017,Ramakrishnan2016} or they assume that a centralized node coordinates the caching decisions~\cite{Llorca2013,Wang2014}, which is unrealistic or difficult to deploy.

In this paper, our goal is to develop a distributed caching policy that preserves the benefits that network coding brings to NDN for the realistic case when the caches have limited capacity. We propose \textit{PopNetCod}, a popularity-based caching policy for network coding enabled NDN architectures. PopNetCod is a caching policy in which routers distributedly estimate the popularity of the content objects based on the received Interest. Based on this information, each router decides which Data packets to insert or evict from its cache. The decision to cache a particular Data packet is taken before the Data packet arrives at the router, \textit{i.e.}, while processing the corresponding Interest. Since the first routers to process Interests in their path to the source are the edge routers, this helps to cache the most popular Data packets closer to the network edges, which reduces the data delivery delay~\cite{Dabirmoghaddam2014,Fayazbakhsh2013,Sun2014}. To avoid caching the same Data packet in multiple routers over the same path, routers communicate the Data packets that they decide to cache by setting a binary flag in the Interests to be forwarded upstream. This increases the Data packet diversity in the caches. When the cache of a router is full and a Data packet should be cached, the router decides which Data packet should be evicted from its cache based on the popularity information.

We implement the proposed caching policy on top of ndnSIM~\cite{ndnSim}, based on the NetCodNDN codebase~\cite{Saltarin2016,Saltarin2017}. We evaluate the performance of PopNetCod in a Netflix-like video streaming scenario, designed using parameters available in the literature~\cite{Netflixblog2015,Boettger2016,NetflixISPindex}. In comparison with a caching policy that uses the NDN's default Leave Copy Everywhere (LCE) placement policy and the Least Recently Used (LRU) eviction policy, PopNetCod achieves a higher cache-hit rate, which translates into higher video quality at the clients and reduced load at the sources.

The remainder of this paper is organized as follows. Section~\ref{sec:related_works} provides an overview of the related works. Section~\ref{sec:system_overview} describes the system architecture. Section~\ref{sec:caching_video} introduces the problem of caching in network coding enabled NDN for data intensive applications. Then, Section~\ref{sec:caching_popularity_driven} presents our caching policy, PopNetCod. A practical implementation of the PopNetCod caching policy is described in Section~\ref{sec:popnetcod_practical}. Section~\ref{sec:popnetcod_evaluation} presents the evaluation of the PopNetCod caching policy.

\section{Related Work}
\label{sec:related_works}

Caching policies are needed to deal with caches that have limited capacity. Caching policies decide which Data packets are placed into the cache (\textit{placement}), as well as which data packets are evicted from the cache when the cache is full and a new Data packet should be cached (\textit{eviction}). There are placement algorithms that consider content popularity to decide which Data packets routers allow in their caches~\cite{Yeh2014,Li2016, Cho2012,Abani2016}.
Specifically, \textit{VIP}~\cite{Yeh2014} is a framework for joint Interest forwarding and Data packet caching. This scheme uses a ``virtual control plane'' that operates on the Interest rate and a ``real plane'' which handles Interests and Data packets. It is shown that the design of joint algorithms for routing and caching is important for NDN. Thus, this scheme proposes distributed control algorithms that operate in the virtual control plane with the aim of increasing the number of Interests satisfied by in-network caches. \textit{PopCaching}~\cite{Li2016} is a popularity-based caching policy in which the popularity is computed online, without the need for a training phase. This makes \textit{PopCaching} robust in dynamic popularity settings. However, \textit{PopCaching} is designed for caching systems with a single cache in the path, while in this paper we are interested in networks of caches. \textit{WAVE}~\cite{Cho2012} is a placement algorithm that determines the number of Data packets that should be cached for a given file with the help of an access counter. The number of Data packets to cache increases exponentially with the value of the access counter. The main idea of \textit{WAVE}, which partially caches a content object according to the local popularity, is also adopted by the caching policy that we propose in this paper. However, \textit{WAVE} does not facilitate edge caching, since the most popular data is cached closer to the source and slowly moves towards the edges as the number of requests increase. \textit{Progressive}~\cite{Abani2016} is another partial caching algorithm, which exploits the content popularity to decide how many Data packets should be cached for each name prefix. The cache placement decision is taken when the Interests are received, which helps to cache the most popular content at the network edge. However, this approach lacks an eviction algorithm, and hence it cannot be deployed when the cache capacity is limited.

None of the approaches above consider the use of network coding~\cite{Ahlswede2000}, and all are evaluated in single-path scenarios. Given the benefits that network coding brings to multipath communications in NDN~\cite{Montpetit2012,Saltarin2016,Saltarin2017,Ramakrishnan2016}, some approaches have been proposed to improve the benefits of caching in network coding enabled NDN architectures~\cite{Llorca2013,Wang2014,Wu2013}. \textit{NCCAM}~\cite{Llorca2013} and \textit{NCCM}~\cite{Wang2014} propose optimal solutions to the problem of efficiently caching in network coding enabled NDN. However, both approaches need a central entity that is aware of the network topology and the Interests, which does not scale well with the number of network nodes. \textit{CodingCache}~\cite{Wu2013} is an eviction policy in which routers, before evicting a Data packet, apply network coding to the Data packet by means of combining it with other Data packets with the same name prefix that will remain in the cache. Due to the increased Data packet diversity in the network, the cache-hit rate is improved. However, in \textit{CodingCache} Interest aggregation and Interest pipelining are problematic, limiting the benefits that network coding brings to the NDN architecture.

\section{Overview of Network Coding Enabled NDN}
\label{sec:system_overview}

\subsection{Data Model}
\label{sec:data_model}

We consider a set of content objects $\mathcal{P}$ that is made available by a \textit{content provider} to a set of \textit{end users}. Each content object is uniquely identified by a name $n$. Clients use this name to request that particular content object. Each content object is divided into a set of Data packets $\mathcal{P}_{n}$, such that the size of each Data packet does not exceed the Maximum Transmission Unit (MTU) of the network. The set of Data packets $\mathcal{P}_{n}$ that compose a content object is divided into smaller sets of Data packets, which are known as \textit{generations}~\cite{Chou2007}. The size of each generation $g$ is a design parameter chosen to enable network coding at scale. The set of Data packets that form the generation $g$ is denoted as $\mathcal{\hat{P}}_{n,g}$ and a network coded Data packet belonging to generation $g$ is represented by $\hat{p}_{n,g}$.

\subsection{Router Model}
\label{sec:ndn}

The routers have three main tables: a \textit{Content Store (CS)}, where they cache Data packets to reply to future Interests, a \textit{Pending Interest Table (PIT)}, where they keep track of the Interests that have been received and forwarded, to know where to send the Data packets backward to the clients, and a \textit{Forwarding Information Base (FIB)}, which associates upstream faces with name prefixes, to route the Interests towards the sources. In order to enable the use of caching policies in the NetCodNDN architecture, we extend its design by adding a new module called Content Store Manager (CSM). The CSM manages the content store by enforcing a determined caching policy.

Whenever a router receives an Interest $\hat{i}_{n,g}$, it first verifies if it can reply to this Interest with the Data packets available in the CS. The router replies to the Interest if it is able to generate a network coded Data packet that has high probability of being innovative when forwarded on the path where the Interest arrived, \textit{i.e.}, if the generated Data packet is linearly independent with respect to all the Data packets that have been sent over the face where the Interest arrived. In this case, the router generates a new Data packet by randomly combining the Data packets in its CS and then sends it downstream over the face where the Interest arrived. Otherwise, the router forwards the Interest to its upstream neighbors to receive a new Data packet that enables it to satisfy this Interest. However, if the router has already forwarded one or multiple Interests with the same name prefix $(n,g)$ and it expects to receive enough Data packets to reply to all the pending Interests stored in the PIT, the router simply \textit{aggregates} this Interest in the PIT, and waits for enough innovative Data packets to arrive before replying to the Interest.

Whenever a router receives a Data packet $\hat{p}_{n,g}$, it first determines if the Data packet is innovative or not. A Data packet $\hat{p}_{n,g}$ is innovative if it is linearly independent with respect to all the Data packets in the CS of the router, \textit{i.e.}, if it increases the rank of $\mathbf{\hat{P}}^{r}_{n,g}$. Non-innovative Data packets are discarded. If the Data packet $\hat{p}_{n,g}$ is innovative, the router sends the Data packet to the CSM, which decides to cache it or not according to the caching policy. Finally, the router generates a new network coded Data packet and sends it over every face that has a pending Interest to be satisfied.

\subsection{Content Store Model}
\label{sec:content_store_model}

The Content Store (CS) is a temporary storage space in which a router $r$ can cache Data packets that it has received and considers useful to reply to future Interests. The maximum number of Data packets that can be cached in the CS is given by $M$, while the set of Data packets that are cached in the CS is denoted as $\mathcal{\hat{P}}^{r}$. Thus, $|\mathcal{\hat{P}}^{r}| \leq M$.

Data packets in the CS are organized in CS entries. Each CS entry contains a set of network coded Data packets, $\mathcal{\hat{P}}^{r}_{n,g}$, that belong to the same generation $g$. Since the CS has a limited capacity of $M$ Data packets, then $\sum_{n,g} |\mathcal{\hat{P}}^{r}_{n,g}| \leq M$. The Data packets that compose a CS entry are stored in a matrix $\mathbf{\hat{P}}^{r}_{n,g}$, where each row is a vector $\mathbf{\hat{p}}_{n,g}$ that represents the network coded Data packet $\hat{p}_{n,g}$.

Router $r$ generates a network coded Data packet $\hat{p}_{n,g}$ by randomly combining the Data packets $\mathbf{\hat{P}}^{r}_{n,g}$ in its CS. Thus, $\hat{p}_{n,g} = \sum_{j=1}^{|\mathbf{\hat{P}}^{r}_{n,g}|} a_{j} \cdot \mathbf{\hat{p}}^{(j)}_{n,g}$, where $a_{j}$ is a randomly selected coding coefficient and $\mathbf{\hat{p}}^{(j)}_{n,g}$ is the $j$th Data packet in $\mathbf{\hat{P}}^{r}_{n,g}$.

Additionally to the matrix $\mathbf{\hat{P}}^{r}_{n,g}$, each CS entry also stores a counter $\sigma^{f}_{n,g}$ for each face $f$ of router $r$. This counter measures the number of Data packets generated by applying network coding to the Data packets stored in matrix $\mathbf{\hat{P}}^{r}_{n,g}$ that have already been sent over face $f$,
\textit{i.e.}, it measures the amount of information from matrix $\mathbf{\hat{P}}^{r}_{n,g}$ that has been transmitted from router $r$ to its neighboring node connected over face $f$. The counter $\sigma^{f}_{n,g}$ is used to compute the number of network coded Data packets with name prefix $(n,g)$ that the router can generate with the Data packets cached in its CS and have high probability of being innovative to its neighboring node connected over face $f$. This number is denoted as $\xi^{f}_{n,g}$ and is computed as follows:
\begin{equation}
  \xi^{f}_{n,g} = \mathtt{rank}(\mathbf{\hat{P}}^{r}_{n,g}) - \sigma^{f}_{n,g} \text{.}
  \label{equ:caching_xi}
\end{equation}

When a Data packet with name prefix $(n,g)$ is evicted from the CS of router $r$, the amount of information in the matrix $\mathbf{\hat{P}}^{r}_{n,g}$ is reduced by 1. Correspondingly, the value of $\sigma^{f}_{n,g}$ is decreased by 1 for all faces.

\section{Caching in Network Coding Enabled NDN}
\label{sec:caching_video}


Whenever a router $r$ receives an Interest $\hat{i}_{n,g}$ over face $f$, it either \textit{(i)} replies with a Data packet $\hat{p}_{n,g}$, if it can generate a network coded Data packet that has high probability of being innovative to its neighboring node connected over face $f$, \textit{i.e.}, $\xi^{f}_{n,g} > 0$, or, otherwise, \textit{(ii)} forwards the Interest $\hat{i}_{n,g}$ upstream.

If at time $t$ router $r$ receives the Interest $\hat{i}_{n,g}$, a cache-hit is defined as:
\begin{equation}
  h^{f}_{n,g}(t) =
  \begin{cases}
      1, & \text{if } \xi^{f}_{n,g} > 0 \\
      0, & \text{otherwise.}
    \end{cases}
  \label{equ:caching_instant_cachehit}
\end{equation}

Let us now assume that during a time period $[t,t+T]$ router $r$ receives a set of Interests $\mathcal{I}(t,T)$. The cache-hit rate during this time period is defined as follows:
\begin{equation}
  H(t,T) = \frac{1}{T} \sum_{t'=t}^{t+T} h^{f}_{n,g}(t') \text{.}
  \label{equ:caching_cache_hit}
\end{equation}

The overall cache-hit rate seen by router $r$ at time $t$ can be computed as follows:
\begin{equation}
  H(t) = \lim_{T \to \infty} H(t,T) = \lim_{T \to \infty} \frac{1}{T} \sum_{t'=t}^{t+T} h^{f}_{n,g}(t') \text{.}
  \label{equ:caching_cache_hit_total}
\end{equation}

To make optimal use of the limited CS capacity, the objective of each router is to maximize the number of Interests that it can satisfy with the Data packets available in its CS, \textit{i.e.}, maximize its overall cache-hit rate.
Achieving a high cache-hit rate at the routers is beneficial for both clients and sources. For the sources, an increased cache-hit rate reduces their processing load and bandwidth needs, since the number of Interests that they receive is reduced. For the clients, the delivery delay is reduced, since the Interests are satisfied with Data packets cached at routers closer to them.

It is clear from~(\ref{equ:caching_instant_cachehit}), (\ref{equ:caching_cache_hit}), and~(\ref{equ:caching_cache_hit_total}) that in order to maximize the overall cache-hit rate, routers should maintain the value of $\xi^{f}_{n,g}$ high enough so that most of the Interests received can be satisfied with the Data packets in their CS. However, since in this paper we consider that the routers' CS have limited capacity, it is unfeasible for a router to cache all the Data packets that it receives~\cite{Montpetit2012,Saltarin2016,Saltarin2017,Ramakrishnan2016}. Optimal solutions to this issue have been proposed in previous works~\cite{Llorca2013,Wang2014}, which consider a central controller that knows the network topology and is aware of all the Interests received by the routers. However, these solutions do not scale well with the size of the network, since they require a high number of signaling messages and a powerful enough controller. Hence, in this work we consider that each router decides online and independently from other routers if a Data packet should be cached or not, and which Data packet should be evicted from the CS when it is full. This is achieved by using a distributed caching policy $\pi$ that maximizes the overall cache-hit rate $H(t)$ of each router,
\begin{equation}
  \max_{\pi} H(t) \text{.}
  \label{equ:caching_cache_hit_max}
\end{equation}

The optimal caching policy $\pi$ predicts which Interests will be received in the future, so that the router caches the Data packets that will be useful to satisfy those Interests.

\section{The PopNetCod Caching Policy}
\label{sec:caching_popularity_driven}

In this section, we present our popularity-based caching policy for network coding enabled NDN, called PopNetCod. To increase the overall cache-hit rate, the PopNetCod caching policy exploits real-time data popularity measurements to determine the number of Data packets that each router should cache for each name prefix. In order to determine which Data packets to cache in and/or evict from the CS, such that the overall cache-hit rate is maximized, PopNetCod performs the following steps. First, it measures the popularity of the different name prefixes contained in the Interests that pass through it. Then, it uses this popularity to predict the Interests that it will receive. Finally, it uses this prediction to determine in an online manner the Data packets that should be cached and the ones that should be evicted from the CS.

\subsection{Popularity Prediction}

\begin{figure}[t]
	\centering
  \includegraphics[width=.4\textwidth]{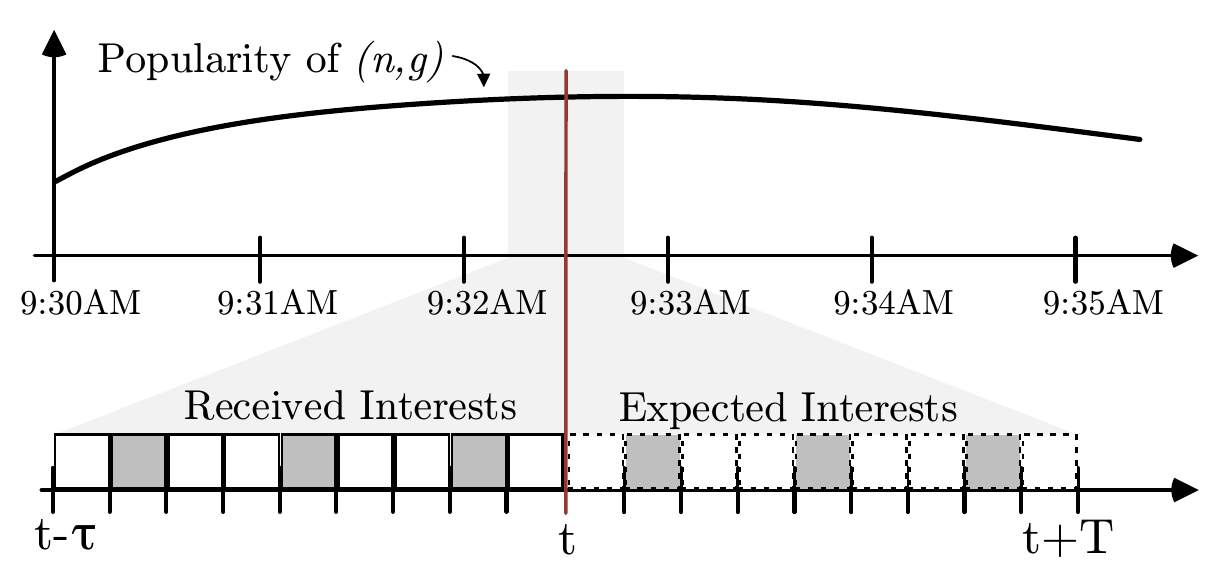}
  \caption{Popularity prediction for the name prefix $(n,g)$.}
  \label{fig:caching_time}
\end{figure}

The popularity prediction in PopNetCod is based on the fact that the rate $\lambda^{f}_{n,g}(t)$ at which Interests for a particular content object arrive at a router $r$ over face $f$ at time $t$ tends to vary smoothly, as shown in Fig.~\ref{fig:caching_time}. Thus, router $r$ can predict the rate of the Interests that it will receive in the near future by observing the Interests that it recently received. Let us denote $\mathcal{I}^{f}_{n,g}(\tau,t)$ as the set of Interests for the name prefix $(n,g)$ that router $r$ has received over face $f$ in the past period $[t-\tau,t]$, where $t$ is the current time and $\tau$ is the observation period.
Let us also denote $\mathcal{I}^{f}(\tau,t)$ as the total set of Interests for all name prefixes received over face $f$ during the period $[t-\tau,t]$. Using the sets $\mathcal{I}^{f}_{n,g}(\tau,t)$ and $\mathcal{I}^{f}(\tau,t)$, router $r$ can compute the average Interest rate for the name prefix $(n,g)$ over face $f$ as follows:
\begin{equation}
  \lambda^{f}_{n,g}(\tau, t) = \frac{|\mathcal{I}^{f}_{n,g}(\tau,t)|}{|\mathcal{I}^{f}(\tau,t)|} \text{,}
  \label{equ:caching_interest_rate}
\end{equation}

Note that since the average Interest rate does not vary abruptly, the average Interest rate $\lambda^{f}_{n,g}(\tau,t)$ of the recent period $[t-\tau,t]$ will be very close to that expected in the near future, \textit{i.e.}, in the period $[t,t+T]$ where $T$ is the length of the prediction period. Thus, $\lambda^{f}_{n,g}(\tau,t) = \lambda^{f}_{n,g}(t,T)$, which hereafter we denote as $\lambda^{f}_{n,g}(t)$. The PopNetCod caching policy uses $\lambda^{f}_{n,g}(t)$ to predict the number of Interests with name prefix $(n,g)$ that will be received over face $f$ in the near future, and hence, to allocate more storage space in the CS to Data packets with higher cache-hit probability.

In order to prepare the CS for the Interests that the router may receive, the PopNetCod caching policy maps the received Interest rate to the capacity of the CS, such that name prefixes with high rate are allocated more space in the CS. The number of network coded Data packets with name prefix $(n,g)$ that the router should cache in its CS at time $t$ to satisfy the Interests expected over face $f$ is denoted as $M^{f}_{n,g}(t)$ and computed as:
\begin{equation}
  M^{f}_{n,g}(t) =
  \begin{cases}
      \lambda^{f}_{n,g}(t) \cdot M, & \text{if } \lambda^{f}_{n,g}(t) \cdot M < |\mathcal{\hat{P}}_{n,g}| \\
      |\mathcal{\hat{P}}_{n,g}| , & \text{otherwise.}
    \end{cases}
  \label{equ:caching_utility_function_2}
\end{equation}

\subsection{PopNetCod Placement}

In the PopNetCod caching policy, the placement decision is taken following the reception of an Interest. Whenever a router decides to cache the Data packet that is expected as a reply to the received Interest, it sets a flag on the Interest signaling upstream routers about its decision. In the case of a set flag, the upstream nodes do not consider this Interest for caching. Since the edge routers (\textit{i.e.}, the routers that are directly connected to the clients) are the first ones that have the possibility to decide whether they will cache a Data packet, the PopNetCod caching policy naturally enables edge caching. This is inline with recent works~\cite{Dabirmoghaddam2014, Fayazbakhsh2013, Sun2014} arguing that most of the gains from caching in NDN networks come from edge caches, and thus, it is natural to cache the most popular content at edge routers.

Whenever a router receives an Interest $\hat{i}_{n,g}$ over face $f_t$ at time $t$, the PopNetCod caching policy follows the next steps to decide if the Data packet $\hat{p}_{n,g}$ should be cached. First, it uses popularity prediction to compute $M^{f}_{n,g}(t)$, \textit{i.e.}, the total number of Data packets that it aims to cache for name prefix $(n,g)$, as defined in (\ref{equ:caching_utility_function_2}). Then, it computes the number of Data packets that it should cache in order to satisfy the expected Interests as:
\begin{equation}
  \delta^{f}_{n,g}(t) = M^{f}_{n,g}(t) - \xi^{f}_{n,g}(t) \ \forall f \in \mathcal{F} \text{.}
  \label{equ:caching_utility_function_3}
\end{equation}

Finally, the caching policy decides to cache the Data packet $\hat{p}_{n,g}$ that is expected as reply to the received Interest if the average number of Data packets needed by all the faces is greater than 0. However, it should be noted that the Data packet $\hat{p}_{n,g}$ will not be useful to the node connected over the downstream face $f_t$ over which the Interest arrived. This is because when the Data packet $\hat{p}_{n,g}$ arrives at the router, it is sent to face $f_t$ in order to satisfy the received Interest. Then, replying with the same Data packet to a subsequent Interest received over the same face $f_t$ does not add any innovative information, \textit{i.e.}, the Data packet is considered as duplicated. Instead, the expected Data packet $\hat{p}_{n,g}$ is potentially useful for all the nodes connected over all the other downstream faces of the router. For this reason, the average number of Data packets needed is measured only over the downstream faces different to the one over which the Interest arrived. It is computed as:
\begin{equation} 
  \Delta^{+}_{n,g}(t) = \frac{1}{|\mathcal{F}^{r}|-1} \sum_{\substack{f \in \mathcal{F}\\f \neq f_t}} \delta^{f}_{n,g}(t) > 0 \text{,}
  \label{equ:caching_placement}
\end{equation}

where $\mathcal{F}^{r}$ denotes the downstream faces of router $r$.

\subsection{PopNetCod Eviction} 


The steps followed by the PopNetCod caching policy to decide how many Data packets with name prefix $(n,g)$ can be evicted from the router's CS are the following. Similarly to the placement case, first, the caching policy uses popularity prediction to compute $M^{f}_{n,g}(t)$, \textit{i.e.}, the number of Data packets that it aims to cache for name prefix $(n,g)$. Then, it computes the number of Data packets that it can evict from its CS and still satisfy the expected Interests as:
\begin{equation}
  \tilde{\delta}^{f}_{n,g}(t) = \mathtt{rank}(\mathbf{\hat{P}}^{r}_{n,g}) - M^{f}_{n,g}(t) \forall f \in \mathcal{F} \text{.}
  \label{equ:caching_eviction_npkt}
\end{equation}

Finally, the number of Data packets the router can evict from a particular name prefix $(n,g)$ is computed as the minimum number of Data packets that it can evict over all the faces:
\begin{equation}
  \Delta^{-}_{n,g}(t) = \min_{f \in \mathcal{F}} \tilde{\delta}^{f}_{n,g}(t) \text{.}
  \label{equ:caching_eviction}
\end{equation}

\section{Practical Implementation of PopNetCod}
\label{sec:popnetcod_practical}

In this section, we describe a practical implementation of the PopNetCod caching policy in the NetCodNDN architecture~\cite{Saltarin2017}.
First, we describe the signaling between routers, which is used to prevent routers of the same path to cache duplicate Data packets. Next, we present the Interest processing algorithm, where placement decisions are made. Finally, we describe the Data packet processing algorithm for placement enforcement, eviction decision, and eviction enforcement.

\subsection{Signaling Between Routers}

The PopNetCod caching policy is distributed and requires very limited signaling between routers. The only signaling that exists between routers to implement the PopNetCod caching policy is a binary flag added to the Interest and Data packets that is used to inform neighbor routers that an expected Data packet will be cached or that a received Data packet has been cached. Distributed caching policy decisions help to keep the complexity of the system low and to make our system scalable to a large number of routers.

Each Interest $\hat{i}_{n,g}$ carries a flag \texttt{CachingDown}, which is set to $1$ by a router when it decides to cache the Data packet $\hat{p}_{n,g}$ that is expected to come as reply to the Interest. This flag informs upstream routers that another router downstream has already decided to cache the Data packet that is expected to come as reply to this Interest. The routers receiving an Interest with the \texttt{CachingDown} flag set to $1$ do not consider to cache the Data packet that is expected to come as reply to this Interest, therefore reducing the number of duplicated Data packets in the path and the processing load in the nodes.

Since Interests for network coded data do not request particular Data packets, but rather any network coded Data packet with the requested name prefix, the routers need a way to know that a Data packet has been already cached by another router, so that they avoid caching duplicated Data packets. For this reason, each Data packet $\hat{p}_{n,g}$ has a flag \texttt{CachedUp}, which is set to $1$ by a router when it caches this Data packet in its CS. This flag informs the downstream routers that another router has already cached this Data packet. A router receiving a network coded Data packet with the flag set to $1$ does not consider it for caching. Instead, it waits for another Data packet with the same name prefix that has not been cached upstream. This ensures that a Data packet is cached by only one router on its way to the client.

\subsection{Status Information at Routers}

Each router implementing the PopNetCod caching policy should store information that assists to identify the Data packets that should be cached or evicted. In particular, the router needs to keep the \textit{Recently received Interests} information to compute the popularity prediction. Moreover, since the placement decision takes place when the Interest is received, the router needs to remember the \textit{Names to be cached}, such that the selected Data packets are cached when they arrive. Finally, since the popularity information can vary over time, the routers should keep a list with the \textit{Names to consider for eviction}, which is used when they decide about eviction. Below, we describe the data structures used to store this information.

\begin{itemize}

  \item \textit{Recently received Interests} --- The router maintains a list $\mathbf{L}^{f}$ for each face $f$ of the router, where it stores the names of the Interests $\mathcal{I}^{f}(\tau,t)$ received over face $f$ during the period $[\tau,t]$. The parameter $\tau$ controls how much into the past is observed by the router to compute the popularity prediction. Together with the name prefix, each element in $\mathbf{L}^{f}$ also stores the time $t_i$ at which the Interest was received, such that it can be removed from $\mathbf{L}^{f}$ at time $t_i+\tau$.

  \item \textit{Names to be cached} --- The router maintains a table $\mathbf{A}$, where it stores the name prefixes (\textit{i.e.}, the content object name appended with the generation ID)  and the number of the Data packets that should be cached. When the router receives an Interest $\hat{i}_{n,g}$ and the PopNetCod caching policy decides that the network coded Data packet that is expected as reply should be cached, the router adds its name prefix $(n,g)$ to the list $\mathbf{A}$. Then, whenever a network coded Data packet arrives, the router looks for the name prefix of the Data packet in the list $\mathbf{A}$. If it finds a match, it caches the Data packet. 

  \item \textit{Names to consider for eviction} --- The router also maintains a queue $\mathbf{E}$, where it stores the name prefixes of the CS entries that can be considered for Data packet eviction. When a name prefix $(n,g)$ is removed from the list $\mathbf{L}^{f}$, the popularity of this name prefix decreases, \textit{i.e.}, it is a good candidate to consider for eviction. Thus, each time a name prefix is removed from $\mathbf{L}^{f}$, it is added to $\mathbf{E}$. 

\end{itemize}

\subsection{Interest Processing}

\begin{figure}[t]
	\centering
  \includegraphics[width=0.37\textwidth]{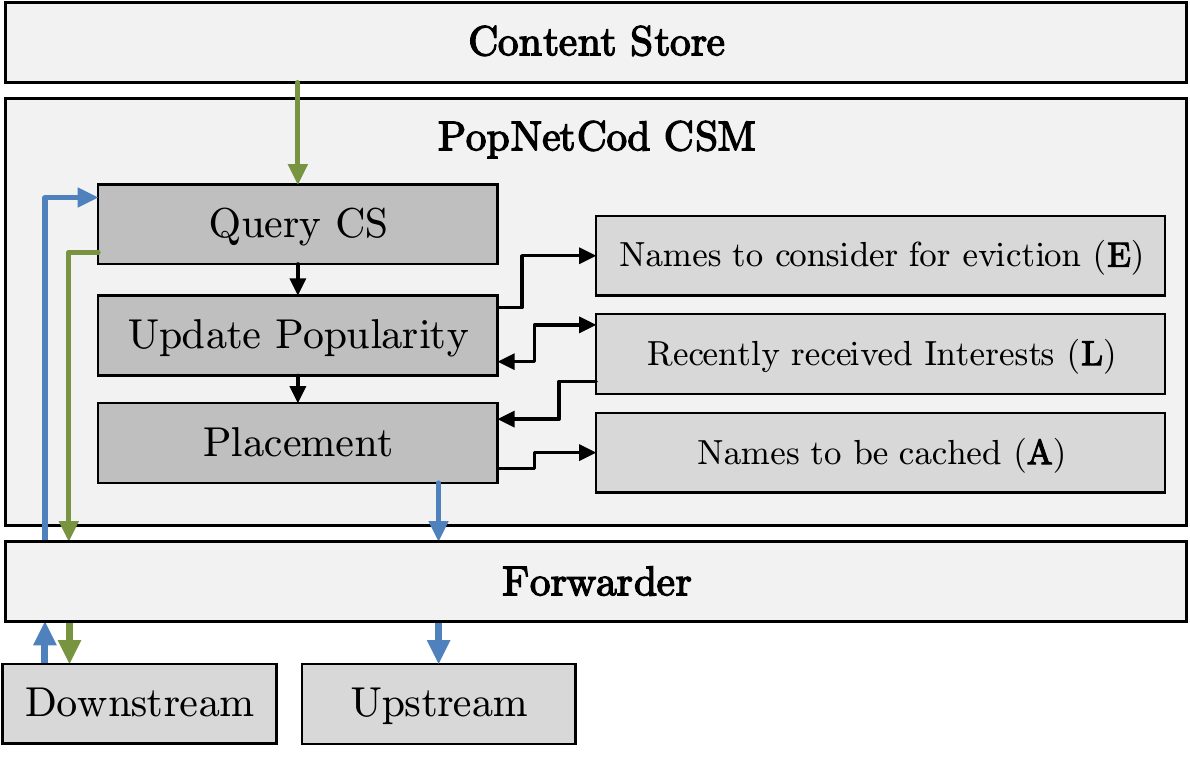}
  \vspace{-8pt}
  \caption{Access to the CS and the Status Information during the Interest processing in a CSM configured with the PopNetCod caching policy.}
  \label{fig:caching_interest_csm}
\end{figure}


\begin{algorithm}[t]
	\caption{Interest processing at the CSM}
	\label{algo:caching_interest_processing}
	\begin{algorithmic}[1]
		\REQUIRE $\hat{i}_{n,g}$, $f$
    \STATE $t \gets$ current time
    \IF {Flag \texttt{CachingDown} in $\hat{i}_{n,g}$ is set to $1$} \label{algo:line:cachingdown_start}
        \IF [$\hat{i}_{n,g}$ can be satisfied from the CS]{$\xi^{f}_{n,g} > 0$}
          \STATE Generate a Data packet $\hat{p}_{n,g}$ from the CS
          \STATE Return $\hat{p}_{n,g}$
        \ELSE
          \STATE Return $\hat{i}_{n,g}$
        \ENDIF \label{algo:line:cachingdown_end}
		\ELSE
				\STATE Add $(n,g)$ to $\mathbf{L}^{f}$ \label{algo:line:add_to_L}
        \IF [$\hat{i}_{n,g}$ can be satisfied from the CS]{$\xi^{f}_{n,g} > 0$} \label{algo:line:serve_from_cs}
            \STATE Generate a Data packet $\hat{p}_{n,g}$ from the CS
            \STATE Return $\hat{p}_{n,g}$ \label{algo:line:serve_from_cs_end}
        \ELSIF {$\hat{i}_{n,g}$ will be aggregated by the PIT}
          \STATE Return $\hat{i}_{n,g}$ \label{algo:line:aggregating}
        \ELSE
          \STATE Update $\mathbf{L}$. \label{algo:line:update_L}
          \COMMENT{Algorithm~\ref{algo:caching_removing_expired}}
          \IF [$\hat{p}_{n,g}$ should be cached]{$\Delta^{+}_{n,g}(t) > 0$} \label{algo:line:should_cache}
            \STATE Insert $(n,g)$ into $\mathbf{A}$
            \STATE Set the flag \texttt{CachingDown} of $\hat{i}_{n,g}$ to $1$
            \STATE Return $\hat{i}_{n,g}$ \label{algo:line:should_cache_end}
          \ELSE
            \STATE Return $\hat{i}_{n,g}$ \label{algo:line:final_line}
          \ENDIF
        \ENDIF
		\ENDIF
	\end{algorithmic}
\end{algorithm}

\begin{algorithm}[t]
	\caption{Update $\mathbf{L}$}
	\label{algo:caching_removing_expired}
	\begin{algorithmic}[1]
    \FORALL {$f \in \mathcal{F}^{r}$}
      \FORALL {expired entries $(n_l,g_l)$ in $\mathbf{L}^{f}$}
          \STATE Remove $(n_l,g_l)$ from $\mathbf{L}^{f}$
          \STATE Add $(n_l,g_l)$ to $\mathbf{E}$
      \ENDFOR
    \ENDFOR
  \end{algorithmic}
\end{algorithm}

As depicted in Fig.~\ref{fig:caching_interest_csm}, when a CSM configured with the PopNetCod caching policy receives an Interest $\hat{i}_{n,g}$ from downstream, it \textit{(i)} determines if the Interest can be replied from the CS. Then, if the CSM could not reply to the Interest with the content of its CS, it \textit{(ii)} updates the popularity information, and, \textit{(iii)} determines if the Data packet that is expected as reply to this Interest should be cached. The CSM should provide the NetCodNDN forwarder with either a Data packet that should be sent as reply to the Interest, or an Interest that should be forwarded upstream. Below we describe the details of this procedure, which is summarized in Algorithm~\ref{algo:caching_interest_processing}.

After receiving an Interest $\hat{i}_{n,g}$, the CSM first checks the flag \texttt{CachingDown} to see if any previous node downstream in the path has decided to cache the Data packet that is expected as reply to this Interest (lines~\ref{algo:line:cachingdown_start} to~\ref{algo:line:cachingdown_end}). If the flag \texttt{CachingDown} is set to $1$, then the CSM only checks its CS to determine if the Interest can be satisfied from the CS.
If this is possible, \textit{i.e.}, if $\xi^{f}_{n,g}$ is greater than $0$, it generates a network coded Data packet from the CS and provides it to the NetCodNDN forwarder, which sends it over face $f$. If the Interest can not be satisfied from the CS, the CSM provides the same Interest to the NetCodNDN forwarder, which forwards it upstream.

If the flag \texttt{CachingDown} is set to $0$, the CSM first inserts name $(n,g)$ of the Interest into the list $\mathbf{L}^{f}$ (line~\ref{algo:line:add_to_L}). Then, the CSM checks if it can satisfy the Interest with the content of the CS (lines~\ref{algo:line:serve_from_cs} to~\ref{algo:line:serve_from_cs_end}).
If this is possible, \textit{i.e.}, if $\xi^{f}_{n,g}$ is greater than $0$, it generates a network coded Data packet from the CS and provides it to the NetCodNDN forwarder which sends it over face $f$. Otherwise, the node needs to forward the Interest to its neighbor nodes. If the router does not send the Interest upstream, but aggregates it in the PIT with a previously received Interest, the CSM does not need to do anything else and provides the Interest to the NetCodNDN forwarder, which aggregates it (line~\ref{algo:line:aggregating}). If the Interest will not be aggregated, then the CSM determines if it will cache the Data packet with name prefix $(n,g)$ that is expected as reply to this Interest, by computing $\Delta^{-}_{n,g}(t)$ using Eq.~(\ref{equ:caching_placement}).

In order to obtain an accurate value of $\Delta^{-}_{n,g}(t)$, the CSM first updates the popularity information, removing all the expired elements from $\mathbf{L}^{f}$ and adding their name prefix to the list $\mathbf{E}$ of name prefixes to be considered for eviction (line~\ref{algo:line:update_L}). This procedure is summarized in Algorithm~\ref{algo:caching_removing_expired}.
Then, the CSM computes the value of $\Delta^{+}_{n,g}(t)$. If $\Delta^{+}_{n,g}(t) > 0$, it means that the Data packet should be cached. In this case, the CSM inserts name prefix $(n,g)$ into the list $\mathbf{A}$, sets the flag \texttt{CachingDown} on the Interest $\hat{i}_{n,g}$ to $1$ and, finally, provides the modified Interest to the NetCodNDN forwarder, which forwards it upstream (lines~\ref{algo:line:should_cache} to~\ref{algo:line:should_cache_end}). If $\Delta^{+}_{n,g}(t) \leq 0$, then the CSM provides the same Interest to the NetCodNDN forwarder, which forwards it upstream (line~\ref{algo:line:final_line}).

\subsection{Data Packet Processing}

\begin{figure}[t]
	\centering
  \includegraphics[width=0.37\textwidth]{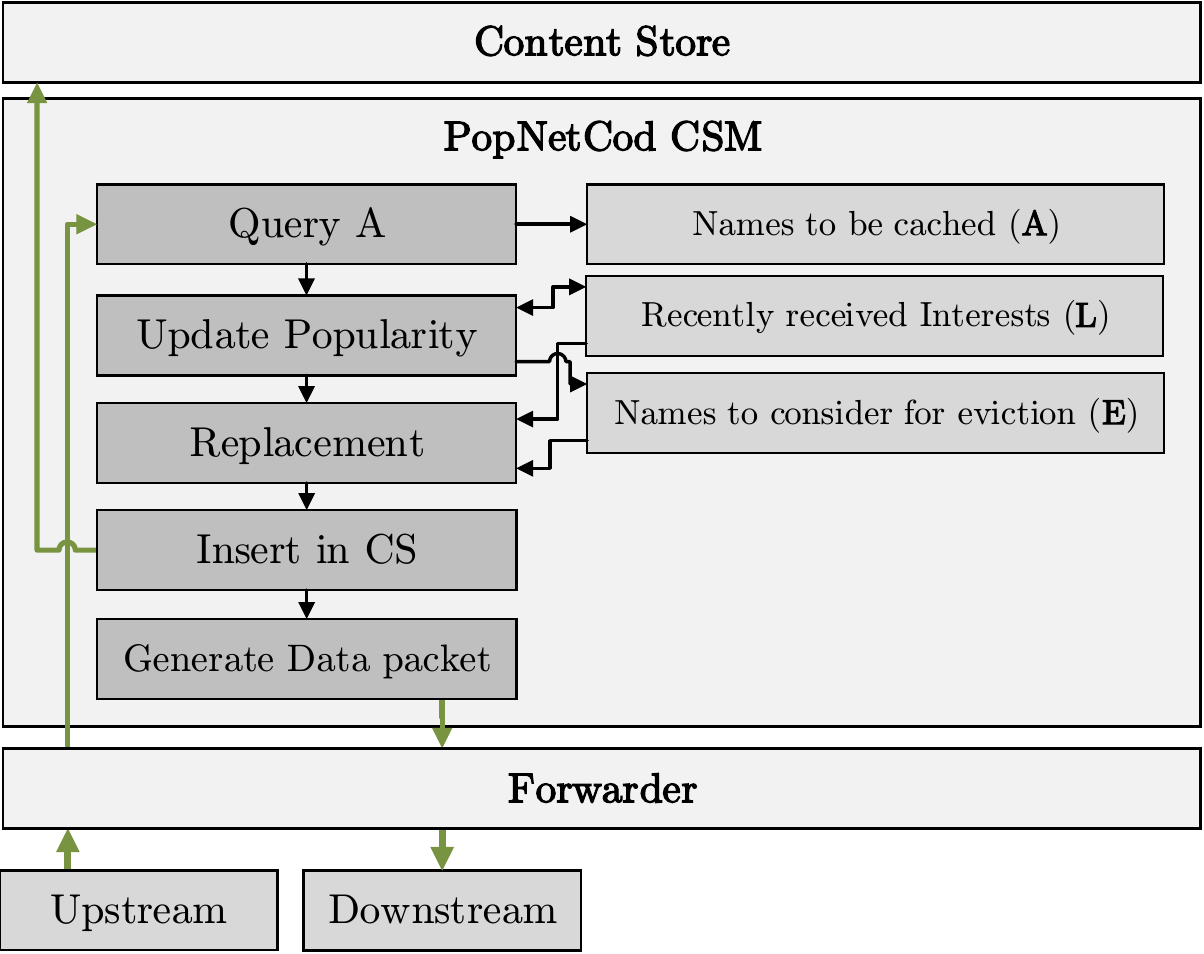}
  \vspace{-3pt}
  \caption{Access to the CS and the Status Information during the Data packet processing in a CSM configured with the PopNetCod caching policy.}
  \label{fig:caching_datapacket_csm}
\end{figure}

\begin{algorithm}[t]
	\caption{Data packet processing at the CSM}
	\label{algo:caching_data_processing}
  \begin{algorithmic}[1]
		\REQUIRE $\hat{p}_{n,g}$
    \IF {Flag \texttt{CachingUp} in $\hat{p}_{n,g}$ is set to $1$} \label{algo:line:cached_up_1}
      \STATE Return $\hat{p}_{n,g}$
    \ELSIF {$(n,g) \notin \mathbf{A}$} \label{algo:line:not_in_A}
      \STATE Return $\hat{p}_{n,g}$
    \ELSE
      \STATE Update $\mathbf{A}$ \label{algo:update_A_data}
      \IF [The CS is full]{$|\mathcal{P}^{r}| == M$} \label{algo:cs_full}
        \STATE Update $\mathbf{L}$ \label{algo:update_L_data}
        \COMMENT{Algorithm~\ref{algo:caching_removing_expired}}
        \WHILE {$|\mathcal{P}^{r}| == M$}
          \STATE Select an element $(n_e,g_e)$ from $\mathbf{E}$
          \IF {$\Delta^{-}_{n_e,g_e}(t) > 0$}
            \STATE Evict $\Delta^{-}_{n_e,g_e}(t)$ Data packets with name prefix $(n_e,g_e)$ from the CS
          \ENDIF
        \ENDWHILE
      \ENDIF \label{algo:cs_full_end}
      \STATE Insert $\hat{p}_{n,g}$ into the CS
      \STATE Generate a Data packet $\hat{p}^{*}_{n,g}$ from the CS
      \STATE Set the flag \texttt{CachingDown} of $\hat{p}^{*}_{n,g}$ to $1$
      \STATE Return $\hat{p}^{*}_{n,g}$
		\ENDIF
	\end{algorithmic}
\end{algorithm}

As depicted in Fig.~\ref{fig:caching_datapacket_csm}, when a CSM configured with the PopNetCod caching policy receives a network coded Data packet $\hat{p}_{n,g}$ from upstream, it \textit{(i)} determines if the Data packet should be cached in the CS, by consulting $\mathbf{A}$. If the Data packet should be cached, the CSM  ensures that there is enough free space in the CS, \textit{(ii)} updating the popularity information and \textit{(iii)} executing the cache replacement procedure if needed. Finally, the CSM \textit{(iv)} inserts the received Data packet into the CS, and \textit{(v)} generates a new network coded Data packet that should be forwarded downstream. This procedure is detailed below and summarized in Algorithm~\ref{algo:caching_data_processing}.

After receiving a Data packet $\hat{p}_{n,g}$, the CSM first checks the flag \texttt{CachedUp} to determine if any router upstream has already cached this Data packet. If the flag \texttt{CachedUp} has been set to $1$, then, the CSM understands that another router upstream has already cached this Data packet. In this case, the CSM returns the Data packet to the NetCodNDN forwarder, which replies to any matching pending Interest (line~\ref{algo:line:cached_up_1}).

When the flag \texttt{CachedUp} is set to $0$, then the CSM first verifies if any entry in $\mathbf{A}$ matches name prefix $(n,g)$. If there is no matching entry, the CSM returns the Data packet to the NetCodNDN forwarder (line~\ref{algo:line:not_in_A}). If there is a match, the Data packet should be cached, and $\mathbf{A}$ is updated by increasing the counter of the matching entry by one (line~\ref{algo:update_A_data}). However, if the CS is full, the CSM first needs to release some space in the CS (lines~\ref{algo:cs_full} to~\ref{algo:cs_full_end}). To evict Data packets, the CSM goes through the list $\mathbf{E}$, each time selecting a name prefix $(n_e,g_e)$ and computing the number of Data packets that can be evicted for the name prefix using Eq.~(\ref{equ:caching_eviction}). If this number is greater than $0$, then the CSM evicts the corresponding number of Data packets from the CS and interrupts the scan of the list.
Note that, since the cached Data packets are network coded, the CSM does not need to decide which particular Data packets from the CS entry $\mathcal{\hat{P}}_{n,g}$ it should evict from the CS, but it can select randomly network coded Data packets from the CS entry and evict them. After evicting at least one Data packet, the CSM caches the received Data packet $\hat{p}_{n,g}$. Then, the router generates a new Data packet $\hat{p}^{*}_{n,g}$ by applying network coding to the cached Data packets with name prefix $(n,g)$. Since the new Data packet $\hat{p}^{*}_{n,g}$ contains the cached Data packet $\hat{p}_{n,g}$, the router sets the flag \texttt{CachedUp} of $\hat{p}^{*}_{n,g}$ to $1$.
Finally, the CSM provides Data packet $\hat{p}^{*}_{n,g}$ to the NetCodNDN forwarder, which uses it to reply to pending Interests with name prefix $(n,g)$.

\section{Evaluation}
\label{sec:popnetcod_evaluation}

In this section, we evaluate the performance of the PopNetCod caching policy in an adaptive video streaming architecture based on NetCodNDN~\cite{Saltarin2017}. First, we describe the evaluation setup. Then, we present the caching policies with which we compare the PopNetCod caching policy. Finally, we show the performance evaluation results.

\subsection{Evaluation Setup}
\label{sec:caching_evaluation_setup}

We consider a layered topology consisting of 1 source, 123 clients, and 45 routers connecting the clients and the sources. The routers are arranged in a two-tier topology, with 10 routers directly connected to the source and 35 edge routers directly connected to the clients. The links connecting the routers between them and the links connecting the routers to the source have a bandwidth of $20Mbps$. The bandwidth of the links connecting the clients to the routers follow a normal distribution, with mean $4Mbps$ and standard deviation $1.5$. These values are chosen based on the Netflix ISP Speed Index~\cite{NetflixISPindex}. Each client is connected with two routers, considering that nowadays most end-user devices have multiple interfaces, \textit{e.g.}, LTE, Wi-Fi.

For the evaluation, we consider that the source offers 5 videos for streaming, each one composed of 50 video segments with a duration of 2 seconds each, \textit{i.e.}, in total, each video has a duration of 100 seconds. The video segments are available in three different representations, $\mathcal{Q} = \{ 480p, ~720p, ~1080p \}$ with bitrates $\{ 1750kbps,\allowbreak~3000kbps, \allowbreak~5800kbps \}$, respectively. These values for the representations and bitrates are according to the values that had been used by Netflix~\cite{Netflixblog2015}. As presented in Section~\ref{sec:data_model}, the content objects (\textit{i.e.}, the video segments in our evaluation scenario) are divided into Data packets and generations, in order to implement network coding.
In particular, for the representations $\mathcal{Q} = \{ 480p, ~720p, ~1080p \}$, each video segment is divided into $\{ 359, 615, 1188 \}$ Data packets of $1250$ bytes each, and $\{ 4, 7, 12 \}$ generations, respectively. Thus, in total, the source stores $540,500$ Data packets. All the routers are equipped with content stores able to cache between $0.9\%$ and $2.3\%$ of the total Data packets available at the source.

The clients randomly choose a video to request and start the adaptive video retrieval process at a random time during the first 5 seconds of the simulation. The network coding operations are performed in a finite field of size $2^8$. The clients use the \textit{dash.js} adaptation logic~\cite{Timmerer2016} to choose the representation that better adapts to the current conditions, \textit{i.e.}, the measured goodput and the number of buffered video segments.

\subsection{Benchmarks}
\label{sec:caching_evaluation_benchmarks}

We compare the performance of our caching algorithm with the following benchmarks:

\begin{itemize}
  \item \textit{LCE-NoLimit} --- The placement policy is Leave Copy Everywhere (LCE). We assume that the CSs of the routers have enough space to store all the videos. 
  \item \textit{LCE+LRU} --- The placement policy is LCE, while the eviction policy is Least Recently Used (LRU), which evicts Data packets with the least recently requested name.
  \item \textit{NoCache} --- In this setting, the routers do not have a CS, \textit{i.e.}, all the Data packets should be retrieved from the source.
\end{itemize}

\subsection{Evaluation Results}
\label{sec:caching_evaluation_results}

\begin{figure*}[t]
  \minipage[t]{0.32\textwidth}
    \input{cache-hit-average.pgf}
    \vspace{-6pt}
    \caption{Average cache-hit rate in the routers.}
    \label{plot:cache-hit-average}
  \endminipage\hfill
  \minipage[t]{0.32\textwidth}
    \input{goodput.pgf}
    \vspace{-6pt}
    \caption{Average goodput perceived by the clients.}
    \label{caching:plot:goodput}
  \endminipage\hfill
  \minipage[t]{0.32\textwidth}
    \input{representation-bandwidth-netcodndn-PopNetCod.pgf}
    \vspace{-6pt}
    \caption{Percentage of video segments delivered in each of the representations, with PopNetCod.}
    \label{caching:plot:representations_popnetcod}
  \endminipage\hfill
\end{figure*}

\begin{figure*}[t]
  \minipage[t]{0.32\textwidth}%
    \input{representation-bandwidth-netcodndn-LCE+LRU.pgf}
    \vspace{-6pt}
    \caption{Percentage of video segments delivered in each of the representations, with LCE+LRU.}
    \label{caching:plot:representations_lce+lru}
  \endminipage\hfill
  \minipage[t]{0.32\textwidth}
    \input{representation-bandwidth-netcodndn-LCE+NoLimit.pgf}
    \vspace{-6pt}
    \caption{Percentage of video segments delivered in each of the representations, with LCE+NoLimit.}
    \label{caching:plot:representations_lce+nolimit}
  \endminipage\hfill
  \minipage[t]{0.32\textwidth}%
    \input{data-from-source.pgf}
    \vspace{-6pt}
    \caption{Load reduction in the source, measured as the percentage of Data packets provided by caches.}
    \label{plot:datafromsource}
  \endminipage\hfill
\end{figure*}

We first evaluate the average cache-hit rate at the routers. In Fig.~\ref{plot:cache-hit-average}, we can see that by using the PopNetCod caching policy, the routers achieve a higher cache-hit rate than with LCE-LRU. This is because with PopNetCod the number of Data packets cached for a certain name prefix increases smoothly, according to the popularity. In comparison, with LCE+LRU all Data packets received by the router are cached, and the least recently used are evicted from the CS when the capacity is exceeded. Thus, if a router receives Data packets that are requested by a single client, the router still caches them, wasting storage capacity that could be used to cache more popular Data packets that are requested by multiple clients. We can also see that the LCE+NoLimit caching policy defines an upper bound to the cache-hit rate at the routers, since caching all the Data packets with unlimited CS capacity represents the best caching scenario. On the contrary, the NoCache case, where the routers do not have CS capacity, defines a lower bound to the cache-hit rate. Note that in our evaluation the NoCache policy has a non-zero cache-hit rate because our measurement of cache-hit rate also includes Interest aggregations, which is what is being measured in this case.

The increased cache-hit rate that the PopNetCod caching policy brings to the routers has two major consequences: \textit{(i)} the goodput at the clients increases, which enables the adaptation logic to choose higher quality representations when bandwidth is sufficient, and \textit{(ii)} the source receives less Interests, meaning that its processing and network load is reduced.

Let us first evaluate the impact that the increased cache-hit rate at the routers has for the clients. In Fig.~\ref{caching:plot:goodput}, it is shown that by using PopNetCod, the clients benefit from an increased goodput, compared to the LCE+LRU policy. This is a consequence not only of the increased cache-hit rate in the network, but also because PopNetCod caches the most popular content in the network edge, which reduces the content retrieval delay. The percentage of video segments delivered to the clients for each of the available representations (\textit{i.e.}, $480p$, $720p$, and $1080p$) with the PopNetCod and LCE+LRU caching policies is shown in Figs.~\ref{caching:plot:representations_popnetcod} and~\ref{caching:plot:representations_lce+lru}, respectively.
We can see that, compared to the LCE+LRU policy, with the PopNetCod caching policy a higher percentage of video segments are delivered in the highest representation available, \textit{i.e.}, $1080p$. This happens because the Data packet retrieval delay is reduced, since more Interests are being satisfied from the routers' content stores, which increases the goodput measured by the clients. The percentage of video segments delivered to the clients in each of the available representations with the upper bound LCE+NoLimit caching policy can be seen in Fig.~\ref{caching:plot:representations_lce+nolimit}.

Finally, we analyze the impact that the increased cache-hit rate in the routers has for the sources by measuring the load reduction at the source. This metric measures the percentage of Data packets received at the clients that have not been directly provided by the source. It is computed as $1 - N^{sent}_{\mathcal{S}} / N^{rcvd}_{\mathcal{C}}$, where $N^{sent}_{\mathcal{S}}$ denotes the total number of Data packets sent by the source, and $N^{rcvd}_{\mathcal{C}}$ denotes the total number of Data packets received by all the clients. In Fig.~\ref{plot:datafromsource}, we can see that by using the PopNetCod caching policy, the source load is reduced by up to 10\% more than by using LCE+LRU, when the CS size is 12.5K Data packets. Note that the load reduction on the source in the NoCache scenario is larger than $0$, even if no Data packet is being served from the CSs. This is because the Interest aggregation at the routers makes it possible to serve multiple Interests with the same Data packet, reducing the number of Data packets delivered by the source.

\section{Conclusions}
\label{sec:conclusion}

In this paper, we have presented PopNetCod, a popularity-based caching policy for data intensive applications communicating over network coding enabled NDN. PopNetCod is a distributed caching policy, where each router aims at increasing its local cache-hit rate, by measuring the popularity of each content object and using it to determine the number of Data packets for each content object that it caches in its content store. PopNetCod takes cache placement decisions when Interests arrive at the routers, which naturally enables edge caching.
The evaluation of the PopNetCod caching policy is performed in a Netflix-like video streaming scenario. The results show that, in comparison with a caching policy that uses the LCE placement policy and the LRU eviction policy, PopNetCod achieves a higher cache-hit rate. The increased cache-hit rate reduces the number of Interests that the source should satisfy, and also increases the goodput seen by the clients. Thus, our caching policy presents benefits for the content providers, by reducing the load of its servers and hence its operative costs, and for the end-users, who are able to watch higher quality videos.

\bibliographystyle{IEEEtran}

\begin{thebibliography}{10}
\providecommand{\url}[1]{#1}
\csname url@samestyle\endcsname
\providecommand{\newblock}{\relax}
\providecommand{\bibinfo}[2]{#2}
\providecommand{\BIBentrySTDinterwordspacing}{\spaceskip=0pt\relax}
\providecommand{\BIBentryALTinterwordstretchfactor}{4}
\providecommand{\BIBentryALTinterwordspacing}{\spaceskip=\fontdimen2\font plus
\BIBentryALTinterwordstretchfactor\fontdimen3\font minus
  \fontdimen4\font\relax}
\providecommand{\BIBforeignlanguage}[2]{{%
\expandafter\ifx\csname l@#1\endcsname\relax
\typeout{** WARNING: IEEEtran.bst: No hyphenation pattern has been}%
\typeout{** loaded for the language `#1'. Using the pattern for}%
\typeout{** the default language instead.}%
\else
\language=\csname l@#1\endcsname
\fi
#2}}
\providecommand{\BIBdecl}{\relax}
\BIBdecl

\bibitem{CiscoVNI}
``{Cisco Visual Networking Index: Forecast and Methodology, 2016-2021},'' White
  Paper, Cisco Systems Inc., Jun. 2016.

\bibitem{Zhang2014}
L.~Zhang, A.~Afanasyev, J.~Burke, V.~Jacobson, K.~Claffy, P.~Crowley,
  C.~Papadopoulos, L.~Wang, and B.~Zhang, ``Named data networking,''
  \emph{SIGCOMM Comp. Comm. Review}, vol.~44, no.~3, pp. 66--73, Jul. 2014.

\bibitem{Jacobson2009}
V.~Jacobson, D.~K. Smetters, J.~D. Thornton, M.~F. Plass, N.~H. Briggs, and
  R.~L. Braynard, ``Networking named content,'' in \emph{Proc. ACM CoNEXT'09},
  Dec. 2009.

\bibitem{Dabirmoghaddam2014}
A.~Dabirmoghaddam, M.~Mirzazad-Barijough, and J.~J. Garcia-Luna-Aceves,
  ``{Understanding Optimal Caching and Opportunistic Caching at the Edge of
  Information-Centric Networks},'' in \emph{Proc. ACM ICN'14}, 2014.

\bibitem{Fayazbakhsh2013}
S.~K. Fayazbakhsh, Y.~Lin, A.~Tootoonchian, A.~Ghodsi, T.~Koponen, B.~Maggs,
  K.~Ng, V.~Sekar, and S.~Shenker, ``Less pain, most of the gain: incrementally
  deployable {ICN},'' in \emph{Proc. ACM SIGCOMM'13}, 2013.

\bibitem{Sun2014}
Y.~Sun, S.~K. Fayaz, Y.~Guo, V.~Sekar, Y.~Jin, M.~A. Kaafar, and S.~Uhlig,
  ``Trace-driven analysis of {ICN} caching algorithms on video-on-demand
  workloads,'' in \emph{Proc. ACM CoNEXT'14}, 2014.

\bibitem{Montpetit2012}
M.-J. Montpetit, C.~Westphal, and D.~Trossen, ``Network coding meets
  information-centric networking: an architectural case for information
  dispersion through native network coding,'' in \emph{Proc. ACM NoM Workshop},
  Jun. 2012.

\bibitem{Saltarin2016}
J.~Saltarin, E.~Bourtsoulatze, N.~Thomos, and T.~Braun, ``{NetCodCCN}: a
  network coding approach for content-centric networks,'' in \emph{Proc. IEEE
  INFOCOM'16}, Apr. 2016.

\bibitem{Ahlswede2000}
R.~Ahlswede, N.~Cai, S.-Y. Li, and R.~Yeung, ``Network information flow,''
  \emph{IEEE Trans. Information Theory}, vol.~46, no.~4, pp. 1204--1216, Jul.
  2000.

\bibitem{Saltarin2017}
J.~Saltarin, E.~Bourtsoulatze, N.~Thomos, and T.~Braun, ``Adaptive video
  streaming with network coding enabled named data networking,'' \emph{IEEE
  Trans. on Multimedia}, vol.~19, no.~10, Oct. 2017.

\bibitem{Ramakrishnan2016}
A.~Ramakrishnan, C.~Westphal, and J.~Saltarin, ``Adaptive video streaming over
  ccn with network coding for seamless mobility,'' in \emph{Proc. IEEE ISM'16},
  Dec. 2016.

\bibitem{Llorca2013}
J.~Llorca, A.~Tulino, K.~Guan, and D.~Kilper, ``Network-coded caching-aided
  multicast for efficient content delivery,'' in \emph{Proc. ICC'13}, 2013.

\bibitem{Wang2014}
J.~Wang, J.~Ren, K.~Lu, J.~Wang, S.~Liu, and C.~Westphal, ``An optimal cache
  management framework for information-centric networks with network coding,''
  in \emph{Proc. IFIP Networking'14}, Jun. 2014, pp. 1--9.

\bibitem{ndnSim}
S.~Mastorakis, A.~Afanasyev, I.~Moiseenko, and L.~Zhang, ``{ndnSIM} 2: an
  updated {NDN} simulator for {NS-3},'' NDN, Tech. Rep.~28, Nov. 2016.

\bibitem{Netflixblog2015}
A.~Aaron, Z.~Li, M.~Manohara, J.~D. Cock, and D.~Ronca, ``The {N}etflix tech
  blog: Per-title encode optimization,''
  https://medium.com/netflix-techblog/per-title-encode-optimization-7e99442b62a2,
  Dec. 2015.

\bibitem{Boettger2016}
T.~B{\"o}ttger, F.~Cuadrado, G.~Tyson, I.~Castro, and S.~Uhlig, ``Open connect
  everywhere: a glimpse at the internet ecosystem through the lens of the
  netflix cdn,'' \emph{arXiv preprint arXiv:1606.05519}, Jun. 2016.

\bibitem{NetflixISPindex}
\BIBentryALTinterwordspacing
``{The Netflix ISP Speed Index},'' {Netflix Inc.}, Dec. 2016. [Online].
  Available: \url{https://ispspeedindex.netflix.com/}
\BIBentrySTDinterwordspacing

\bibitem{Yeh2014}
E.~Yeh, T.~Ho, Y.~Cui, M.~Burd, R.~Liu, and D.~Leong, ``{VIP: A Framework for
  Joint Dynamic Forwarding and Caching in Named Data Networks},'' in
  \emph{Proc. ACM ICN'14}, 2014.

\bibitem{Li2016}
S.~Li, J.~Xu, M.~van~der Schaar, and W.~Li, ``Popularity-driven content
  caching,'' in \emph{Proc. IEEE INFOCOM'16}, Apr. 2016.

\bibitem{Cho2012}
K.~Cho, M.~Lee, K.~Park, T.~T. Kwon, Y.~Choi, and S.~Pack, ``{WAVE:
  Popularity-based and collaborative in-network caching for content-oriented
  networks},'' in \emph{Proc. IEEE INFOCOM'13 Workshops}, Mar. 2012.

\bibitem{Abani2016}
N.~Abani, G.~Farhadi, A.~Ito, and M.~Gerla, ``Popularity-based partial caching
  for information centric networks,'' in \emph{Proc. MedHocNet'16}, 2016.

\bibitem{Wu2013}
Q.~Wu, Z.~Li, and G.~Xie, ``Coding{C}ache: multipath-aware {CCN} cache with
  network coding,'' in \emph{Proc. ACM ICN'13 Workshop}, Aug. 2013.

\bibitem{Chou2007}
P.~Chou and Y.~Wu, ``Network coding for the {I}nternet and wireless networks,''
  \emph{IEEE Sig. Proc. Mag.}, vol.~24, no.~5, pp. 77--85, Sep. 2007.

\bibitem{Timmerer2016}
C.~Timmerer, M.~Maiero, and B.~Rainer, ``Which adaptation logic? {An} objective
  and subjective performance evaluation of http-based adaptive media streaming
  systems,'' \emph{arXiv preprint arXiv:1606.00341}, Jun. 2016.

\end{thebibliography}

\end{document}